# Fluctuations, Correlation and Representative Elementary Volume (REV) in Granular Materials

## P. Evesque

Lab MSSMat, UMR 8579 CNRS, Ecole Centrale Paris
92295 CHATENAY-MALABRY, France, e-mail evesque@mssmat.ecp.fr

**Abstract:**
*In general, the mechanics of granular matter is described using continuum mechanics approach; this requires to introduce the concepts of stress and strain, which are averaged quantities, so that this needs also to introduce the notion of representative elementary volume (REV) above which averaged quantities have some physical meaning. As local quantities fluctuate spatially in granular matter; a local measure of stress and strain shall exhibit fluctuations too, whose typical amplitude depends on the sampling size* L. *This paper discusses this problem and the causes for large scale correlation. Mean stress* $\underline{\sigma}$ *applied to a plane surface of size* $L^2$ *is calculated and its fluctuation amplitude* $\delta\sigma$ *is found when local forces are not correlated; it is found that* $\delta\sigma/\underline{\sigma} \propto 1/L$ . *It is shown also that large scale fluctuations of stress can always be interpreted as an inhomogeneous stress field and that static equilibrium modifies the mean stress applied to a rod (in 2d), even if it does not perturb the contact force distribution. This last result is compared to experiment; which indicates that the number* N *of contacts per rod (in 2d) is* 2<N<3 .

**Pacs # :** 5.40 ; 45.70 ; 62.20 ; 83.70.Fn
________________________________________________________________________________

The engineering approach to granular materials is based on the continuum mechanics formulated at a macroscopic level. In fact, instead of taking into account individual grains, the approach consists in moving to a coarser scale of description by introducing the notion of the Representative Elementary Volume (REV). At this scale, the granular material is considered to be a continuous homogeneous medium. The size of this REV should be large enough with respect to the individual grain size in order to define overall quantities such as stresses and strains, but it should be small enough in order not to hide macroscopic heterogeneity. The aim of this paper is to discuss this notion of REV.

This notion of REV does not seem difficult within this classical continuum mechanics framework. However, it shall be improved to incorporate the existence of " natural " fluctuations of the mechanical properties : let us consider for instance the simplest cases of classical materials; they are governed by thermodynamics, which implies existence of time and space fluctuations at a temperature different from 0 K, because this implies that the entropy is finite which imposes in turn that fluctuations exist at a local scale and that their relative amplitude shall decrease with the size of the volume; so this previous definition of REV can be convenient only in perfectly homogeneous materials at 0 K.

This leads to define the REV as the minimum volume from which one can define the macroscopic properties of the material, taking into account the existence of local fluctuations from REV to REV. These local fluctuations shall obey classical rules of thermodynamics ; in particular, one expects that macroscopic physical quantities obey the central limit theorem, so that relative fluctuations decrease normally with the sample size as any Gaussian distribution.

According to this constraint, the REV will be the minimum volume, whose characteristics fluctuate in an uncorrelated manner and from which one is able to describe the macroscopic quantities and their fluctuations from its distribution characteristics. In other words, let us call $\xi$ the typical size of this REV, and let us consider that the physics of the system can be characterised by a set of physical intensive quantities $\{\cdots,\alpha_i,\cdots\}$; $\{\cdots,\alpha_i,\cdots\}$ are random variables which vary with the spatial position; for sake of simplicity we will consider the case when the fluctuations of $\alpha_i$ and $\alpha_j$ are not correlated for $i \neq j$; then let us define $\{\cdots,\alpha_i(\xi),\cdots\}$ the values of $\{\cdots,\alpha_i,\cdots\}$ measured on a system of size $\xi$, or in other words after averaging in a given sample of size $\xi$; each of these quantities fluctuates from sample to sample and is characterised by its probability distribution which we characterise by its mean $\underline{\alpha}_i(\xi)$ and its width $\delta\alpha_i(\xi)$. If one considers now a sample of size extension L, with L>>$\xi$, one shall expect that the set of physical quantities $\{\cdots,\alpha_i(L),\cdots\}$ measured at this length scale L be a set of random variables characterised by their means $\{\cdots,\underline{\alpha}_i(L),\cdots\}$ and their widths $\{\cdots,\delta\alpha_i(L),\cdots\}$ given by

$$\underline{\alpha}_i(L) >= \underline{\alpha}_i(\xi) \tag{1a}$$

$$\delta\alpha_i(L)/\underline{\alpha}_i(L) = \{\delta\alpha_i(\xi)/\underline{\alpha}_i(\xi)\} \, (\xi/L)^{-m/2} \tag{1b}$$

where m characterises the dimensionality of the space where the average is taken: m=2 if the volume of averaging is a surface ; or m=3 when averaging is taken over a volume. Relations (1a & 1b) stem from the fact that the $\alpha_i$ are random intensive variables which fluctuate in an uncorrelated way for length scales larger than $\xi$ (see ref. [1] for a more complete physical discussion). If they were extensive variables, Eq. (1b) would remain exact [1], but Eq. (1a) would be:

$$\underline{\alpha}_i(L) = [L/\xi]^m \, \underline{\alpha}_i(\xi) \tag{1c}$$

It is worth noting that Eq. (1a-c) works even when the sampling size L/$\xi$ is small so that very little events are taken into account, as far as the events obey Poisson distribution and are not correlated [1].

At last, for length scales smaller than $\xi$, fluctuations of $\alpha_i$ are correlated, so that Eq. (1a-c) are no more valid. In such a case one needs to introduce a correlation function if one is interested byin the variations of the fluctuation amplitude within the volume size [1].

**Application to stress in granular material:**

Consider a granular material made of rigid spheres of equal size (2r); these grains are in contact and submitted to an ensemble of forces. Consider now a plane (P) (cf. Fig. 1), defined by its normal **n**, which cuts this material; it separates the material into two parts, the "left" hand part from the "right" hand one; however, it separates the grains into three categories: those which are entirely in the left zone, those in the right one, and those which are cut by the plane. Nevertheless, one can divide the third series of cut grains into two parts, those which pertain to the left (right) zone; they are these grains whose centres pertain to the left (right) area this plane (see Fig. 1); and one can integrate these grains in the two previous "left" and "right" zones.

In order to estimate the component of the stress tensor in the direction **n**, i.e. **s**·**n**, one can try to replace the action of the left zone by a series of forces distributed on the plane (P); this series of forces shall replace the action of the forces acting on contacts pertaining at the same time to the two species of grains, i.e. left and right grains. However, since the distribution of forces is applied now on the plane (P) and not directly on the grains, one has also to introduce a distribution of torques in order to force the static equilibrium of the grains, since equilibrium imposes sum of torques and sum of forces acting on each grain to be equal to zero. Nevertheless, it is expected that summation of torques shall be zero in mean because each torque depends on the distance of the contact to the plane (P) and since this distribution shall be symmetric.

Anyway, in order to simplify the modelling one can consider a rough surface instead of a flat plane (P); the simplest way to build this rough surface ($\Sigma$) from the plane (P) is to link the contact points which have to be considered using the shortest possible way. These contacts points are the contact points in the vicinity of (P) which pertains to both the left and the right species at the same time. It results from this i) that the surface ($\Sigma$) has the same orientation as (P) in mean, so that it is defined by the same vector **n**, ii) that it surface scales as $(\lambda/r)^2$, where $\lambda$ is the typical length measured on (P) and r is the grain radius.

So, be { ,$\alpha$, } the set of contact points between the grains pertaining at the same time to the two species, i.e. to grains of the two left and right zones. Be **F**$^\alpha$ the force at contact $\alpha$ and **l**$^\alpha$ = 2r **n**$^\alpha$ the vector linking the two centers; be **n** the average direction perpendicular to the plane (P), and **n**$^\alpha$ the unit vector defining the contact surface at contact $\alpha$. In this case, labelling $F_i^\alpha$ the i$^{th}$ component of the force exerted by the left grain on the right grain at contact $\alpha$, the stress tensor **s**, which we define as the average force per unit of surface area is given by (cf. Fig. 1).

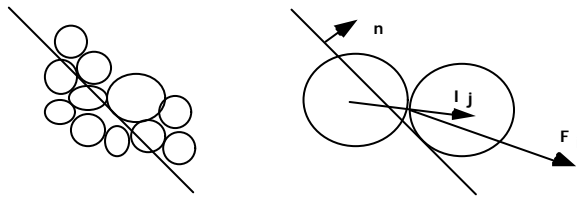

**Figure 1**

$$\mathbf{s}\,\mathbf{n} = \Sigma_j \sigma_{ij} n_j = \Sigma_{\text{contacts } \alpha \text{ to be considered}} \mathbf{F}_i^\alpha \qquad (2)$$

If one considers a very large surface (P) of area $\Sigma$ (i.e. $\Sigma \to \infty$); in this case, one expects that all kinds of contacts are achieved and that effects of correlation are negligible. So, Eq. (2) can be transformed and expressed as a function of (i) the probability $\rho(\mathbf{n}^\alpha)\,d\omega\,d^2n$ of finding a contact of direction $\mathbf{n}^\alpha$ in the volume $d\omega$ and (ii) the conditional probability $p(\mathbf{n}^\alpha,\mathbf{F})\,d^3F$ of finding a force $\mathbf{F}$ at contact $\alpha$ if the direction of contact surface is $\mathbf{n}^\alpha$ within $d^2n$: Labelling $d\mathbf{S}$ the element of surface area on (P), $\mathbf{l}^\alpha$ the vector linking the centres of the two grains belonging to the contact $\alpha$, and defining $\mathbf{n}^\alpha$ as $\mathbf{l}^\alpha = 2r\,\mathbf{n}^\alpha$, one gets that $\mathbf{n}^\alpha$ is also the normal to the surface of contact at contact $\alpha$ since the grains are spheres. So, one can write Eq. (2) as:

$$\int \mathbf{s}.\mathbf{n}\,d\mathbf{S} = \int \rho(\mathbf{n}^\alpha)\,\mathbf{F}\,(2r\,\Sigma\,\mathbf{n}^\alpha.\mathbf{n})\,p(\mathbf{n}^\alpha,\mathbf{F})\,d^2n^\alpha\,d^3F \qquad (3)$$

The term $(2r\,\Sigma\,\mathbf{n}^\alpha.\mathbf{n})$ is introduced in Eq. (3) since the sampling volume defined by (P) depends on the direction $\mathbf{n}^\alpha$ since the distance between contact $\alpha$ and (P) shall not be further than $(\pm r\,\mathbf{n}^\alpha.\mathbf{n})$, leading to $\int d\omega_\alpha = 2r\,\mathbf{n}^\alpha.\mathbf{n}\,\Sigma$.

• *Mean behaviour:*
Let us first consider a large (i.e. $\Sigma \to \infty$) surface and calculate the average stress. So, one can use Eq. (3) and integrate it first on $\mathbf{F}$; denoting $\underline{\mathbf{F}}_{n\alpha}$ the mean value of the contact force when the direction of contact is $\mathbf{n}^\alpha$, one gets:

$$\underline{\mathbf{s}}.\mathbf{n} = \int (2r\,\mathbf{n}^\alpha.\mathbf{n})\,\underline{\mathbf{F}}_{n\alpha}\,\rho(\mathbf{n}^\alpha)\,d\mathbf{n}^\alpha \qquad (4)$$

where $\underline{\mathbf{s}}$ is the average of the local stress tensor, and where $d\mathbf{n}^\alpha = d^2n^\alpha$. One can note that Eqs. (3) and (4) are equivalent to the classical expression for $\underline{\mathbf{s}}$ as given in [2-5]. Furthermore, the mean surface $S_n$ of (P) required to get a single contact of any direction $\mathbf{n}^\alpha$ is:

$$S_n = 1/\left\{\int (2r\,\mathbf{n}^\alpha.\mathbf{n})\,\rho(\mathbf{n}^\alpha)\,d\mathbf{n}^\alpha\right\} \qquad (5)$$

• *Fluctuations:*
So, $\underline{\mathbf{s}}$ is the expected average value, which is measured with an infinite surface; if one uses a smaller surface $\Sigma$ of direction $\mathbf{n}$, the measured value will fluctuate since $\sigma_{\Sigma n}$ depends on the local position; however, if one repeats the measurement in a lot of different points and take the mean, one gets the average given by Eq. (4) too. The fluctuations of $\sigma_{\Sigma n}$ are characterised by the second moment of the distribution, which we label $\delta\sigma_{\Sigma n}$. Calling $S_{o,n}$ the minimum size of the surface, above which no correlation of forces occur, then applying Eq. (1b) one gets:

$$\delta\sigma_{\Sigma n} = (S_{o,n}/\Sigma_n)^{1/2}\,\delta\sigma_{S_{o,n}} \qquad (6)$$

*Case without correlation:*
In order to estimate the amplitude of these fluctuations, one can start considering that forces are not correlated, so that contact force F is a random variable which varies around a mean; obviously the mean itself $\underline{F}_{n\alpha}$ depends on the contact direction $n^a$, as it is already written in Eq. (4); the distribution of $F_{n\alpha}$ is also characterised by its second moment $\delta^2(F_{n\alpha})$, which may depend on the direction $n^a$ too. So, in this case

- $n^a$ fluctuates in direction
- for a given $n^a$, the force **F** fluctuates in direction and intensity;

Let us now consider a large surface, of size $\Sigma$, it applies a given force $F_s$ on the granular material; as it contains a large number of different contacts, this force is the sum of random variables; be $N_\alpha d^2n^\alpha$ the exact number of contacts having the direction $n^a$ within $d^2n^\alpha$, and $\underline{N}_\alpha d^2n^\alpha$ its mean; as it is a random Poisson process, this number fluctuates from surface to surface and its second moment is $\delta^2(N^\alpha) = \underline{N}_\alpha d^2n^\alpha$ ; the forces exerted by this category of contact is a random variable which is the sum of $N_\alpha d^2n^\alpha$ forces which have the same mean $\underline{F}_{n\alpha}$ and same second moment $\delta^2(F_{n\alpha})$. It results from this that these contacts contribute to the total force $F_s$

$$\underline{F}_{n\alpha} \underline{N}_\alpha d^2n^\alpha \qquad \text{in mean and} \qquad (7a)$$

$$\delta^2(F_{n\alpha}) \underline{N}_\alpha d^2n^\alpha \qquad \text{as a standard square deviation} \qquad (7b)$$

$$\text{with} \qquad \underline{N}_\alpha d^2n^\alpha = (2r \Sigma\, n^a.n)\, \rho(n^a)\, d^2n^\alpha \qquad (7c)$$

In the same way, summation over the different categories (i.e. orientations) of contacts leads to evaluate

$$\underline{F}_{s\,n} = \int \underline{F}_{n\alpha}\, \underline{N}_\alpha\, d^2n^\alpha \quad \text{in mean and} \qquad (8a)$$

$$\delta^2(F_{s\,n}) = \int \delta^2(F_{n\alpha})\, \underline{N}_\alpha\, d^2n^\alpha \qquad \text{as a standard square deviation} \qquad (8b)$$

One remarks that Eq. (8a) is equivalent to Eq. (5) as expected, and that Eq. (8b) leads to the mean square deviation:

$$\delta^2(F_{s\,n}) = \Sigma \int \delta^2(F_{n\alpha})\, (2r\, n^a.n)\, \rho(n^a)\, d^2n^\alpha \qquad (9)$$

This Eq. (9) is equivalent to Eq. (6), leading to a standard deviation:

$$\delta\sigma_{\Sigma n} = \delta(F_{s\,n})/\Sigma = (\Sigma)^{-1/2}\, \{\int \delta^2(F_{n\alpha})\, (2r\, n^a.n)\, \rho(n^a)\, d^2n^\alpha\}^{1/2} \qquad (10)$$

which defines $\{S_{o,n}\}^{1/2} \delta\sigma_{So,n}$.

It is worth noting that $\rho(n^\alpha)$, $\underline{F}_{n\alpha}$ and $\delta^2(F_{n\alpha})$ has not always the spherical symmetry, so that $\underline{F}_{s\,n}$ and $\delta\sigma_{\Sigma n}$ do depend on direction **n**. This is the general case when deformation has destroyed the initial isotropic distribution of contacts and forces and when stress is not isotropic.

The mathematical treatment of cases where correlation exists does not differ from the previous one; however the result depends strongly on the kind of correlation. So we will limit here to describe of few different kinds of possible correlation and discuss there effects in the next section.

• *Analysis of the causes of fluctuation correlation:*
Different kinds of correlation can occur in granular matter, which can be classified using the form of Eq. (4): it shows that physical cases separate into cases where correlation exist between i) the direction of contact surfaces, ii) the direction and amplitude of mean contact forces, iii) the local density and iv) the local structure.

• *Steric correlation:* the first kind of correlation is a steric one: Owing to the form of Eqs. (2-4), correlation between directions of force shall exist for distances smaller than r or 2r, since orientations of forces are partly driven by the direction of the surface of the contact $n^\alpha$, and since the directions of 2 contact surfaces $\mathbf{n}^a$ and $\mathbf{n}^{a'}$ pertaining to the same sphere are correlated and depend on the distance between the two contacts, as shown in Fig. 2. This correlation can probably propagate to distance equal to few (1 or 2) grain diameters in disordered packings; however it can propagate on much larger scales when packings are 2d and ordered (Fig. 3-left).

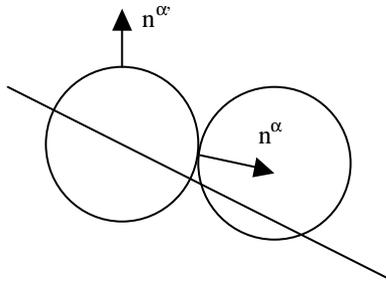

**Figure 2:** i) correlation exists between the direction of contacts $n^\alpha$ & $n^{\alpha'}$ and the distance $r_{\alpha\alpha'}$ between the two contacts $\alpha$ & $\alpha'$. This correlation is obvious inside a given sphere; it exists also for adjacent grains.

• *2d regular packing effects:* In 2d, packings of identical grains are quite often ordered; in this case the grains form a local crystalline structure of well defined symmetry and orientation of principal axis. Most of the time the structure obtained is the triangular one, since it is the densest packing both at the local scale and at the macroscopic scale. This induces a long range correlation of the direction $\mathbf{n}^a$ of contacts; however, at large scale, long range decorrelation of $\mathbf{n}^a$ can occur generated by different mechanisms: i) a change of orientation of the principal axis of the lattice can be induced by a change of orientation of boundary or by a curved boundary or by the deformation process itself ; ii) a change of the lattice structure (from square to triangular) can be imposed by the deformation process and mechanical instability (see Fig. 3). In general, such decorrelation process involves large length scale, and the change of lattice structure or of inclination can be seen as occurring at a grain boundary, if one takes the analogy with what occurs in classic polycrystalline systems.

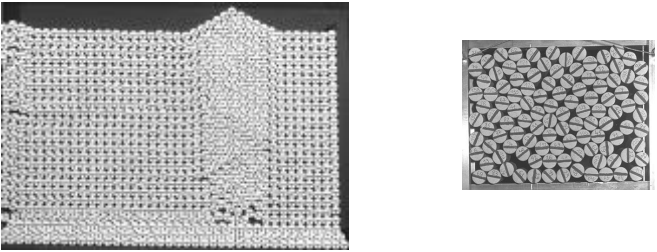

**Figure 3:** Different cases of large length scale correlation of topological arrangements.
  *Left:* this packing was generated from a square lattice after horizontal deformation; it is made of a combination of a square lattice and of a triangular one, with defined orientations which are imposed by boundary conditions and initial state; it looks like a phase transformation.
  *Right:* disordered packing which is made of "bands of square lattice structure" plus some "point-like" defects dispersed in a rather triangular structure; the typical correlation length of the packing structure is 6-to-10 grains about.

Nevertheless, when crystalline zones are large, averaging $\underline{F}_\alpha$ of contact forces is possible for each discrete set of contact direction. Furthermore, as continuity of stress shall be insured when crossing a zone boundary, this implies the existence of some relationships $g_i\{…, \underline{F}_a(phaseA), …,…, \underline{F}_j(phaseB),…\}=0$ between the different mean contact forces of two adjacent phases.

• *Correlation induced by existence of local force paths:* An other cause of correlation is the existence of force paths which spread over the medium as it was demonstrated first by Dantu [6, 7] and confirmed by other workers [8-10]. At first sight, this structure is locally inhomogeneous, since one observes that the series of stressed grains formed a connected network, which contain series of holes made of unstressed zones. This connected structure has a typical size of few grains, which lets think that the correlation is few grain diameters. It is most likely, even if it is not demonstrated yet, that forces are not correlated at larger scale; this comes from the assumption that a good stress tensor can be defined at large scale (see one section further).

Such stress paths introduce a correlation between forces at small length scale in Eq. (3).

• *correlation induced by static equilibrium:* As a matter of fact static equilibrium generates short and long range correlation effect:
- at short range, owing to the fact that each grain is in equilibrium, the sum of the contact forces and of torques shall be 0 on each grain. This reduces the number of free independent forces $N_{Fi}$ on each grain i to the number $N_{ci}$ -2 of contacts on this grain minus 2. This changes the distribution compared to the non correlated case.

For instance, let us consider the experiment performed in [11] on a 2d packing of rods submitted to an average 2d stress, for which one wants to study the statistics of contact forces. The way it was proceeded consisted in measuring the force needed to push each rod parallel to its axis, z (i.e. z is perpendicular to the applied macroscopic 2d stress field). So, the experimental conditions imposed that the projection in the xy plane of the considered contact forces have a direction parallel to the normal of contact, since this contact is sliding in direction z; so the maximum force of friction which can be mobilised on each grain in the z direction is proportional to the sum of the modulus of the projections on the xy plane of the considered contact forces, the coefficient of proportionality being the friction coefficient. Now, owing to the fact that each grain is in equilibrium, the sum of the contact forces and of torques shall be 0 on each grain. This reduces the number of free independent forces $N_{Fi}$ on grain i to the number $N_{ci}$ -1 instead of $N_{ci}$. This influences the shape of the distribution at small forces since this distribution is given by:

$$P(F_{friction}) \, dF_{friction} = \int_{\text{all cases ensuring } F=\Sigma_i F_i} dF_1 \ldots dF_{Ni} \, p(F_1) \ldots p(F_{Ni}) \qquad (11)$$

In the limit of small forces and since $p(F_{contact}) = \exp(-F/F_o) = $ cste when $F \to 0$ [3], this equation behaves as:

$$P(F_{friction}) \, dF_{friction} \propto (F_{contact})^{N_i-2} \, dF_{contact} \quad \text{when } F \to 0 \qquad (12)$$

where $N_i$ is the number of contact of grain i and where $F_{friction}$ and $F_{contact}$ {$P(F_{friction})$, $p(F_{contact})$} are respectively the modulus of the friction force and the contact force {and their probability distribution}. So taking into account the experimental curve reported in [11] for $P(F_{friction})$, which starts rather vertically at small force, one can conclude that the mean number of contacts in a 2d sample is smaller than 3 and larger than 2 in the case of little stressed rods. Indeed, this seems in agreement with what simulation gives in the case of 2d packing with friction [12]. Indeed, a calculation which does not take account of force correlation would lead to $P(F_{friction}) \propto (F_{contact})^{N_i-1}$ when $F \to 0$, leading to a number of contact smaller than 2, which is impossible.

It is worth noting that the reasoning of last paragraph jumped over some difficulty which is discussed now on: the exact number of contacts is an integer for each grain. So, $N_i-2$ shall be an integer in Eq. (12); and Eq. (12) holds for the series of grains which has the same number of contact; hence Eq. (12) calculates $P_{Ni}(F_{friction})$. To get the exact distribution $\underline{P}(F_{friction})$, one has to average on the distribution of $N_i$; as $N_i$ varies from 2 to 6 in the case of rods of equal diameter, with a distribution $W(N_i)$, one expects that $P(F_{friction})$ has a polynomial form, if $W(N_i)$ does not depend on $F_{friction}$:

$$P(F_{friction}) \propto \Sigma_{i=2 \text{ to } 6} \, W(N_i) \, (F_{contact})^{N_i-2} \qquad (13)$$

Eq. (13) is hard to reduce to $(F_{contact})^{N_i-2}$ as in Eq. (12), if $W(Ni)$ is not a function of $F_{friction}$. Under these conditions, it seems that result [11] implies that the contact distribution and the mean number of contact per grain is a function of the local applied force $F_{friction}$. Conversely, this seems to indicate that elasticity may play some part in the contact distribution. However, this modelling shall be improved before complete conclusion can be settled.

- An other question which is raised by the existence of static equilibrium is the following: Does it generate large scale effects? As a matter of fact, it is not possible to conclude surely at the present time. However, one can believe that such effects shall not be so strong since they exist also for classical crystals, poly-crystals, glasses and other solids where they are never considered (to the best of my knowledge). For instance, if one considers a large particle in a liquid, one knows that it is submitted to a Brownian motion which is generated by the fluctuations of local stress; this is due to the fact that static equilibrium is not ensured in liquids. Turning now to the case of an inclusion in a solid, this particle is immovable because static equilibrium is ensured; however, this never implies that the stress field exhibits no spatial fluctuation, but it implies just that the fluctuations are spatially correlated to ensure equilibrium. The case of inhomogeneous stress field of a glassy material is a good example. This is discussed in the next subsections.

- However, before starting this discussion, it is worth emphasising that the stress which is considered in Eqs. (3), (6) & (10) concern the force exerted on a surface; (i) this mean that these results are not affected by correlation due to local equilibrium; (ii) an other consequence is that the noise on **s** scales as $1/L$. Such results are general when all forces acting on the medium are contact forces. However, more care shall be taken when external forces have long range action; this is the case for instance when electric, magnetic and/or gravitational forces have to be considered.

- *Large length-scale heterogeneity of the distribution of contact forces: non homogeneous stress field*:

Consider a macroscopic volume of size $L^2h$, such as $L>>D=2r$ (see Fig.4).

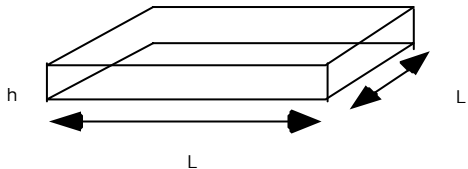

**Figure 4:** Typical volume used to measure the stress $\sigma$; $L>>D=2r$ and $h<<L$, with D the diameter of grains.

Static equilibrium imposes:

$$\sigma_{zz}(z)\, L^2 = \sigma_{zz}(z+h)\, L^2 - 2(Lh)\{\sigma_{zy}(z)+ \sigma_{zx}\} \tag{14}$$

or

$$\sigma_{zz}(z) = \sigma_{zz}(z+h) - 2(h/L)\{\sigma_{zy}(z) + \sigma_{zx}(z)\} \qquad (15)$$

So, Eq. (15) imposes that the difference between the stress tensor $\sigma_{zz}(z+h) - \sigma_{zz}(z)$ at large h, i.e. h>>D, shall always overpass the natural fluctuation amplitude given by Eq. (1c or 6), i.e. $\delta\sigma_{zz} \approx \sigma_{zz}/L$ if $\sigma_{xz} \neq 0$. This means that variations of $\sigma_{zz}(z)$ larger than the typical fluctuations can be interpreted as linked to stress field variations.

To conclude the present discussion, it means that large length scale anomalous fluctuations of the contact force distribution can always be interpreted as linked to the heterogeneity of the stress field. This assumption is assumed implicitly quite often [13].

• *"stress-strain" correlation:* As the mechanical energy delivered to a granular packing which deforms is the product of stress by strain, it is obvious that stress-strain correlation is directly related to a physical quantity. However, this is by far not so simple at a microscopic level as it will be discussed now.

At a local scale, it is clear that the dissipation is due to the sliding of grains in contact. And the dissipation is the summation over all contacts of the scalar product of the contact force by the sliding vector. Contacts which do not slide do not dissipate. So if one knows the sliding field and the contact force distribution one can estimate the dissipation.

On the other hand, the force distribution is related to the stress tensor, as it was studied previously; but there is in principle **no way to relate the strain tensor to the sliding field**, because it is known from plasticity theory that the two fields are incompatible, since the first one ensures continuity of the space and the other one assumes just the contrary. Furthermore, this problem can not be solved as it is done in other cases, such as crystals,…. In order to show that, the next paragraph recalls first briefly the way it is proceeded in case of elastic material forced into plastic deformation.

- *Plastic theory of crystals:* When plasticity theory is applied to elasto-plastic materials, it is done within the model of plastic deformation of crystals, for which plastic deformations are generated via the generation and propagation of dislocations; these dislocations are well defined and the number of different processes is very little. It results from this that the energy of generation and evolution of each process can be calculated, at least approximately, and its interaction with the "elastic" phase is well defined. So knowing the temperature of the material and the stress field and the stress increment, one can use statistical mechanics to predict the process of plastic deformation. So, the evolution of the material shall obey to classical thermodynamics. Hence, this imposes the existence of some equilibrium between the plastic deformation process and the elastic one; and the existence of some stress-strain relation which is called the constitutive law.

- *In the case of granular matter*, the existence of such a law is not obvious since elastic deformation appears to be quite small and since it is difficult to link the stress evolution to the sliding field and the strain field. Nevertheless, the existence of such a stress-strain law seems to exist, because a series of finite-element codes do exist, which are all based on this assumption (i.e. existence of a rheological law) and which predict relatively correctly classical mechanical behaviours. However each code has its own formulation of the law, which turn out to be intricate; this is not a problem, since they all describe the same set of experimental data; it means just that the laws used are too sophisticated.

It is worth mentioning that a much simpler formulation of this law was proposed recently [14,15]; its signature is the existence of the stress-dilatancy law, known as the Rowe's law. It can be shown [13,14] that this relation is equivalent to introduce a dissipation function which depends on stress and deformation. At last, it is worth mentioning that the way Rowe derived his law seems not correct [14,15] .

I have no physical explanation, based on a microscopic model, for the existence of such a constitutive stress-strain relationship. However, it seems likely to be related to the existence of a statistical treatment of the sliding field, so that the sliding field should obey at the same time to some maximum-entropy principle and to some optimum-dissipation principle, which would allow to link stress and strain increments.

If this hypothesis turns out right, the maximum entropy principle should result in the existence of a homogeneous deformation process till the mechanical system remains stable, since maximum of entropy shall require that the number of different deformation process shall increase. However, when the mechanical system becomes unstable localisation of deformation occurs. But all this is quite hypothetical.

## References

[1] L. Landau & E. Lifchitz, *Physique Théorique 5: Physique Statistique*, eds. Mir, Moscou, 1967), pp. 434-437

[2] L. Landau & E. Lifchitz, *Physique Théorique 7: Théorie de l'élasticité*, eds. Mir, Moscou, 1990), pp. 12-18; K. Kantani, "A theory of contact force distribution in granular materials", *Powder Technology* **28**, 167-172, (1981); J. Christoffersen, M. Mehrabadi & S. Nemat-Nasser, "A micro-mechanical description of granular material behaviour", J. Appl. Mech. 48, 339-344, (1981)

[3] P. Evesque, " Statistical mechanics of granular media:An approach à la Boltzmann ", *poudres & grains* **9**, 13-16, (1999)

[4] L. Rothenburg & R.J. Barthurst, "Micro-mechanical features of granular assemblies with planar elliptical particles", *Geotechnique* **42**, 79-95 (1992)

[5] R.J. Barthurst & L. Rothenburg, "Micro-mechanical aspects of isotropic granular assemblies", *J. Appl. Mech Trans. ASME* **55**, 17-23, (1988)

[6] P. Dantu, "Contribution à l'étude mécanique et géométrique des milieux pulvérulents", in *Proc. 4$^{th}$ Int. Conf. Soil Mech. And Foundations Engineering* (Butterworths Scientific Publications, 1957)

[7] P. Dantu , "Etude statistique des forces intergranulaires dans un milieu pulvérulant", *Géotechnique* **18**, 50-55, (1968)

[8] A Drescher & G. de Josselin de Jong, "Photo-elastic verification of a mechanical model for the flow of granular material", J. Mech. Phys. Solids 20, 337-351, (1972)

[9] F. Radjai, "Dynamique des rotations et frottement collectif dans les systèmes granulaires", thèse de l'Université Paris-Sud, (7 décembre 19995)

[10] F. Radjai, D. Wolf, S. Roux, M. Jean & J.J. Moreau, "Force network in dense granular media in *Powders & Grains 93*, eds R.P. Behringer &J.T. Jenkins, (Balkema, Rotterdam, 1997); F. Radjai, D. Wolf, M. Jean & J.J. Moreau, "Bimodal character of stress transmission in granular packing", *Phys. Rev. Lett.* **90**, 61, (1998)

[11] M. Gherbi, R. Gourvès & F. Oudjehane, "Distribution of the contact forces inside a granular material", in *Powders & Grains 93*, ed. C. Thornton, (Balkema, Rotterdam, 1993)

[12] F. Radjai, private communication.

[13] J. P. Bouchaud, M.E. Cates & P. Claudin, J. Phys. I France **5**, 639, (1995)

[14] P. Evesque & C. Stéfani, *J. de Physique II France* **1**, 1337-47 (1991); *C.R. Acad. Sci. Paris* **312**, série II, 581-84 (1991)

[15] P. Evesque, "A Simple Incremental Modelling of Granular-Media Mechanics ", *poudres & grains* **9**, 1-12, (1999)

The electronic arXiv.org version of this paper has been settled during a stay at the Kavli Institute of Theoretical Physics of the University of California at Santa Barbara (KITP-UCSB), in june 2005, supported in part by the National Science Fundation under Grant n° PHY99-07949.

*Poudres & Grains* can be found at :
http://www.mssmat.ecp.fr/rubrique.php3?id_rubrique=402